\documentclass[12pt,preprint]{aastex}

\def\sun{\odot}
\def\earth{\oplus}
\usepackage{epsfig}
\begin{document}
\title{Formation of close in Super-Earths \& Mini-Neptunes: Required Disk Masses \& Their Implications}
 \author{Hilke E. Schlichting\altaffilmark{1} }
 \altaffiltext{1} {Massachusetts Institute of Technology, 77 Massachusetts Avenue, Cambridge, MA 02139-4307, USA} \email{hilke@mit.edu}

\begin{abstract} 
Recent observations by the {\it Kepler} space telescope have led to the discovery of more than 4000 exoplanet candidates consisting of many systems with Earth- to Neptune-sized objects that reside well inside the orbit of Mercury, around their respective host stars. How and where these close-in planets formed is one of the major unanswered questions in planet formation. Here we calculate the required disk masses for {\it in situ} formation of the {\it Kepler} planets. We find that, if close-in planets formed as {\it isolation masses}, then standard gas-to-dust ratios yield corresponding gas disks that are gravitationally unstable for a significant fraction of systems, ruling out such a scenario. We show that the maximum width of a planet's accretion region in the absence of any migration is $2 v_{esc}/\Omega$, where $v_{esc}$ is the escape velocity of the planet and $\Omega$ the Keplerian frequency and use it to calculate the required disk masses for {\it in situ} formation with giant impacts. Even with giant impacts, formation without migration requires disk surface densities in solids at semi-major axes less than 0.1~AU of $10^3-10^5 \rm{~g~cm^{-2}}$ implying typical enhancements above the minimum-mass solar nebular (MMSN) by at least a factor of 20. Corresponding gas disks are below, but not far from, the gravitational stability limit. In contrast, formation beyond a few AU is consistent with MMSN disk masses. This suggests that migration of either solids or fully assembled planets is likely to have played a major role in the formation of close-in super-Earths and mini-Neptunes.
\end{abstract}

\keywords {planets and satellites: dynamical evolution and stability ---protoplanetary disks--- planets and satellites: formation}

\section{Introduction}
NASA's {\it Kepler} mission has been a great success. To date it has discovered over 4000 exoplanet candidates \citep{B13}. The results from the {\it Kepler} mission have provided us, for the first time, with a robust determination of the relative abundances of different-sized planets ranging from Earth-sized bodies all the way to Jupiter-sized planets with periods of less than 100 days. We now know that planets smaller than Neptune are ubiquitous and that about 50\% of all Sun-like stars harbor an exoplanet smaller than Neptune with a period less than 100 days \citep{H12,F13}. The results from {\it Kepler} reveal a new population of planets that consists of Earth- to Neptune-sized bodies that reside well inside the orbit of Mercury around their respective host stars. This new class of planets is unlike anything found in our own Solar System raising fundamental questions concerning their nature and formation. 

Planet formation is generally considered to consist of several distinct stages \citep[e.g.][]{GLS04}. In the first phase, dust settles to the mid-plane of the solar nebula and accumulates into planetesimals \citep{GW73,YS02}. In the second stage, runaway growth leads to the rapid formation of a small number of large, roughly lunar-sized protoplanets \citep[e.g.][]{S72,WS89,SS11}. In the third stage, the growth transitions to oligarchic growth once protoplanets become massive enough to dominate the gravitational stirring in their respective feeding zones \citep[e.g.][]{KI98,R2003}. By the end of oligarchic growth, protoplanets have consumed most of the material in their respective feeding zones and thereby reached their {\it isolation masses}. In the outer parts of the disk, {\it isolation masses} are comparable to the masses of Uranus and Neptune. However, in the inner regions, {\it isolation masses} are only a fraction of an Earth mass. The terrestrial planets are therefore thought to have undergone an additional stage in the planet formation process consisting of collisions of a few dozen protoplanets, called giant impacts \citep{CW98,ACL99}. Numerical modeling of this final stage of terrestrial planet formation \citep{C01} generally produces about the right masses and number of terrestrial planets. The typical eccentricities of those planets are significantly larger than those of the terrestrial planets in our Solar System today, but dynamical friction provided by small, leftover planetesimals \citep{RQ06,SWY12} can dampen the eccentricities to observed values.

What makes the many planetary candidates discovered by {\it Kepler} so intriguing is that they have orbital distances well inside our terrestrial planet region, but their typical sizes, densities and inferred compositions more closely resemble those of Uranus and Neptune \citep{LF13,R14}. Understanding how these close-in planets formed is one of the major unanswered questions in planet formation. 

\citet{BL14} performed detailed numerical simulations of gas accretion onto {\it isolation masses} at formation locations from 0.5 to 4~AU and concluded that the {\it Kepler}-11 systems likely formed further out in the disk with subsequent inward migration. \citet{CL13} proposed recently that close-in super-Earths could have formed {\it in situ} from typical disks that are enhanced by about a factor of 5 compared to the minimum mass solar nebula (MMSN) \citep{H81} and find a radial disk mass surface density profile $\Sigma \propto a^{-1.6}$, which has a similar scaling to the MMSN. However, \citet{RC14} used known {\it Kepler} systems that contained at least 3 planets to construct a MMSN and find that it is inconsistent to assume a universal disk density profile and that many of the resulting disk profiles cannot be explained by viscous gas disk models \citep{CG97,LP74}. \citet{HM12} proposed that 50-100 $M_{\earth}$ of rocky material was delivered to the inner regions of the protoplanetary disk and that the final assembly of planets occurred locally via giant impacts. Finally, \citet{BF13} and \citet{CT14} suggested that inward drifting material is stopped and collected in a pressure maximum in the disk and that planet formation proceeds from there either by core accretion or by gravitational instability.Õ

In this letter we examine the minimum disk masses required for {\it in situ} formation of close-in super-Earths and mini-Neptunes in the absence of migration of solids and/or planets. We calculate the minimum disk masses needed to form these planets as {\it isolation masses} similar to Uranus and Neptune, as assumed by \citet{RB11} and \citet{BL14}, and also determine disk masses required if planets formed with a final stage of giant impacts analogous to the terrestrial planets in the Solar System as suggested by \citet{CL13}. Assuming standard dust-to-gas ratios we examine the stability of the inferred gas disk against gravitational collapse. 

This letter is structured as follows. In section 2.1, we first derive the maximum planet masses that a body can grow to in the absence of migration and use this to infer the local disk surface densities that would have been required for {\it in situ} formation. We show in section 2.2 that, for standard gas-to-dust ratios, a significant fraction of these gas disks are close to, or even beyond, the gravitational stability limit and compare the required disk masses for {\it in situ} formation to the MMSN in section 2.3. Our discussions and conclusions follow in section 3.

\section{Formation of Close In Super-Earths \& Mini-Neptunes}

\subsection{Maximum Planet Masses without Migration}
The largest mass a planet or protoplanet of radius $R$ can grow to in the absence of any migration is its {\it isolation mass}, $M$, defined as the sum of all the material in itÕs local feeding zone, and is given by 
\begin{equation}\label{e1}
M = 2\pi a \Delta a \Sigma
\end{equation}
where $a$ is the semi-major axis, $\Delta a$ the width of the feeding zone and $\Sigma$ the mass surface density of solids in the disk. The width of the feeding zone is given by the radial extent over which the planet can accrete material and therefore depends on the planet's and planetesimals' velocity dispersions. It is usually assumed that both have random velocities less than the Hill velocity, such that their relative velocities are dominated by the Keplerian shear of the disk. In this case $\Delta a \sim 2 v_H/\Omega$, where $\Omega$ is the Keplerian frequency, $v_H=a\Omega (M/3M_{\sun})^{1/3}$ is the planet's Hill velocity, $M_{\sun}$ is the mass of the host star, and the factor of 2 accounts for the contributions from planetesimals residing interior and exterior with respect to the planet. Numerical integrations find that the largest impact parameters leading to accretion are about a factor of 2.5 times larger than the above estimate for the width of the planet's feeding zone \citep{GL90}. This yields an {\it isolation mass} of 
\begin{equation}\label{e0}
M=\frac{(10 \pi \Sigma a^2)^{3/2}}{(3 M_{\sun})^{1/2}}.
\end{equation}
Evaluating the {\it isolation mass} assuming that $\Sigma$ is given by the MMSN, $\Sigma_{MMSN}=7 \times (a/1\rm{AU})^{-3/2}~\rm{g~cm^{-2}}$ \citep{H81}, yields $M \simeq 0.03~M_{\earth}$ at 1AU. Due to these small {\it isolation masses}, the terrestrial planets are believed to have formed from a series of giant impacts of a few dozen protoplanets \citep[e.g.][]{ACL99,C01}. 

\begin{figure}[htp]
\centerline{
\epsfig{file=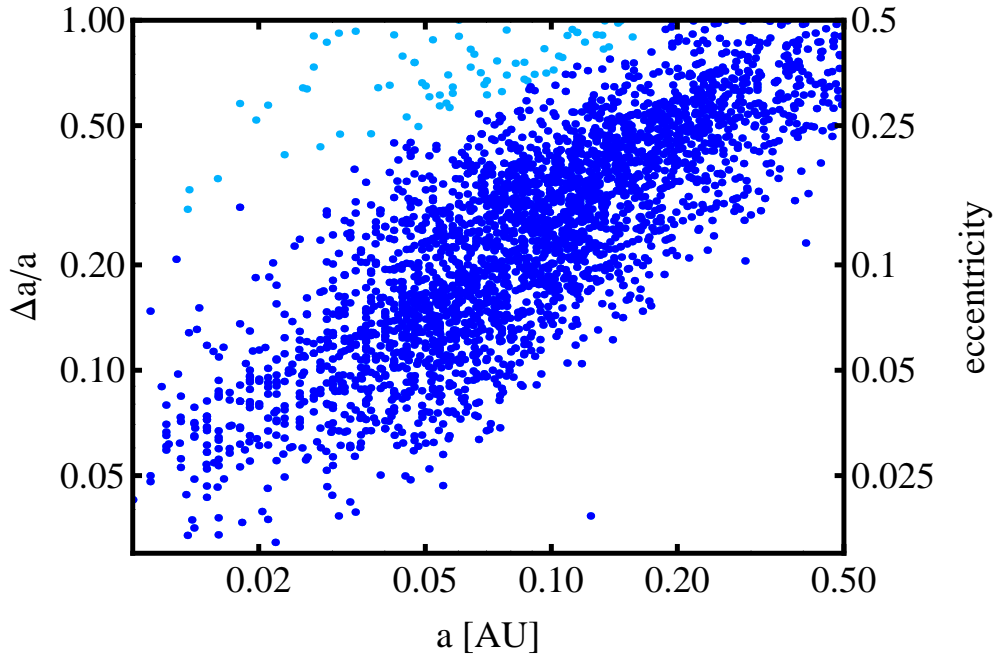, scale=1.3}}
\caption{Maximum width of the accretion zone, $\Delta a$, divided by semi-major axis, $a$, as a function of $a$ for {\it Kepler} planetary candidates. The dark blue points correspond to systems with planetary radii $R \leq 5 R_{\earth}$ and the light blue points to systems with planetary radii $R> 5R_{\earth}$. A density of $2~\rm{g~cm^{-3}}$ was assumed when converting planetary radii into masses throughout this paper. At small distances from the star, the accretion zones are only a small fraction of the planet's semi-major axis. The ratio $\Delta a/a$ can also be thought of as the planet formation efficiency, because its inverse gives an estimate of the number of similar sized planets that should have formed interior to the observed {\it Kepler} planet if the disk extended inwards toward the central star. The y-axis on the right side displays the corresponding eccentricities.}
\label{fig3}
\end{figure}

Using Equation (\ref{e1}) we can also calculate the largest planetary masses that form as a result of giant impacts. Viscous stirring increases the velocity dispersion, $v$, of all bodies in the disk by converting energy associated with the Keplerian sheer into random kinetic energy of the protoplanets. This way protoplanets can mutually stir themselves to a velocity dispersion comparable to their own escape velocity, $v_{esc}$. Once velocity dispersions of $v_{esc}$ are achieved, the collision rate, $\mathcal R_{coll} \simeq n v \pi R^2 (1+(v_{esc}/v)^2)$, exceeds the rate for gravitational stirring, $v^{-1} dv/dt \simeq n v \pi R^2 (v_{esc}/v)^4$, where $n$ is the number density of protoplanets \citep{S72,GLS04}, and $v$ can only be increased significantly further in a single interaction by encounters with minimum encounter distances of less than the protoplanet's radius. Such encounters, however, result in a collision rather than a gravitational deflection. 
Therefore, the maximum distance from which planetesimals and comparably sized protoplanets can be accreted is given by $\Delta a \simeq 2 v_{esc}/\Omega$, which yields
\begin{equation}\label{e3}
\Delta a \simeq 2^{3/2} a \left(\frac{a}{R} \frac{M}{M_{\sun}}\right)^{1/2}.
\end{equation}
This corresponds to eccentricities of $e \simeq \Delta a/2a \simeq \left(2aM/RM_{\sun}\right)^{1/2}$. Figure \ref{fig3} shows the maximum width of the accretion zone, $\Delta a$, divided by the semi-major axis, $a$, as a function of $a$ for {\it Kepler} planetary candidates. At small distances from the star, the accretion zones are only a small fraction of the planet's semi-major axis, which is very different from the assumption made by  \citet{CL13}, who used $\Delta a \sim a$, and requires eccentricities of order unity. The ratio $\Delta a/a$ can also be thought of as the planet formation efficiency, because its inverse gives an estimate of the number of similar sized plants that should have formed interior to the observed {\it Kepler} planet if the disk extended inwards toward the central star. Given the large number of single planet systems discovered by {\it Kepler} \citep{B13}, Figure \ref{fig3} therefore also shows that true {\it in situ} formation must have been very inefficient at small semi-major axis. If most {\it Kepler} candidates formed {\it in situ} at $a < 0.1$~AU without migration then less than 20\% of all solids present were converted into planets observed today. Our results on the maximum accretion width and its implication for the number of planets formed by giant impacts are also the likely explanation for why numerical simulations of {\it in situ} assembly by giant impacts find much fewer single planet systems \citep{HM13} than discovered by {\it Kepler}.

Substituting for $\Delta a$ from Equation (\ref{e3}) into Equation (\ref{e1}) yields a maximum planet mass of
\begin{equation}\label{e2}
M_{max} \simeq \frac{\left[2^{5/2} \pi a^2 \Sigma (\rho/\rho_{\sun})^{1/6} (a/R_{\sun})^{1/2}\right]^{3/2}}{M_{\sun}^{1/2}}.
\end{equation}
The maximum mass in Equation (\ref{e2}) should be close to the absolute maximum mass that a planet can grow to due to giant impacts, because even if the velocity dispersion of the protoplanets could somehow be significantly excited above $v_{esc}$, mutual giant impacts of protoplanets with a random velocities equal to $v_{esc}$ and larger, typically do not lead to accretion \citep{A10}. Evaluating Equation (\ref{e2}) for the MMSN at 1~AU yields $M_{max} \simeq 1.4 M_{\earth}$.
We somewhat overestimate the actual width of the accretion zone because we assume that all the random velocity is excited in the plane rather than distributed in comparable amounts between eccentricity and inclination \citep{IM92}. The actual accretion width will therefore be, on average, smaller by up to a factor of 2 compared to Equation (\ref{e3}). This is also consistent with the typical eccentricities that are found in N-body simulations at the end of giant impacts, which have characteristic values of less than 0.2 \citep{C01}.

\begin{figure}[htp]
\centerline{
\epsfig{file=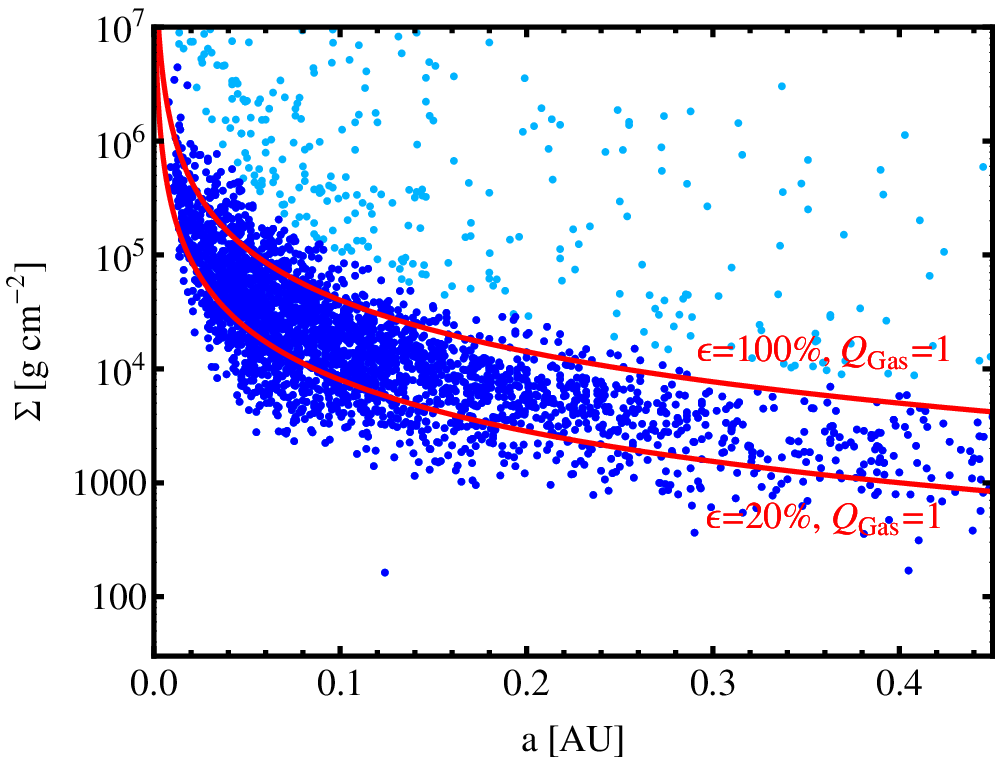, scale=1.3}}
\caption{Mass surface density of solids, $\Sigma$, needed to form the {\it Kepler} candidates as {\it isolation masses}, by accreting all the material in their respective feeding zones, without migration of solids and/or planets. The dark blue points correspond to systems with planetary radii $R \leq 5 R_{\earth}$ and the light blue points to systems with planetary radii $R>5 R_{\earth}$. The upper and lower solid red lines corresponds to the Toomre $Q$ stability parameter of 1 for the corresponding gas disk, assuming a gas-to-dust ratio of 200 and a planet formation efficiency of $\epsilon=$100\% and $\epsilon=$20\%, respectively. A significant fraction of systems fall above the $\epsilon=100\%, Q_{Gas}=1$ line, implying that these disks would be gravitationally unstable to collapse.}
\label{fig4}
\end{figure} 

\begin{figure}[htp]
\centerline{
\epsfig{file=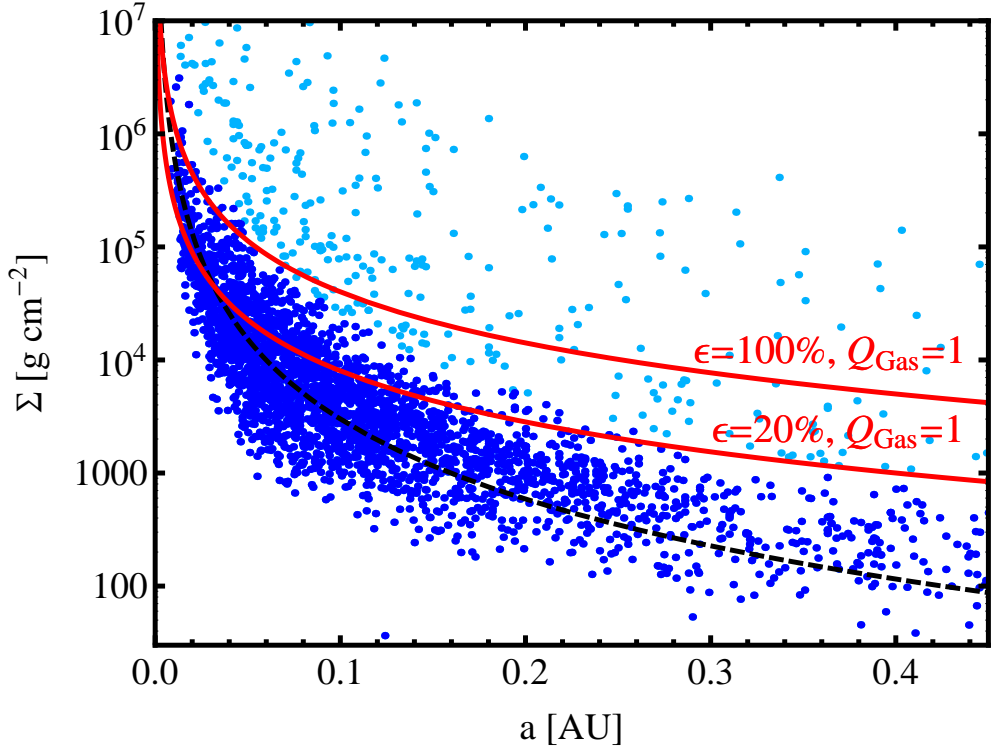, scale=1.3}}
\caption{Same as Figure \ref{fig4} but for solid mass surface density, $\Sigma$, needed to form the {\it Kepler} candidates {\it in situ} with a phase of giant impacts.
The mass surface densities displayed here are calculated assuming $\Delta a \simeq 2v_{esc}/\Omega$. This corresponds to the maximum accretion widths that can result in disks in which protoplanets stir themselves gravitationally. Furthermore, even if the velocity dispersion could be excited significantly above $v_{esc}$, the resulting giant impacts typically would not lead to accretion and may, in some cases, result in erosion instead \citep{A10}. The dashed black line is the best fit disk surface density model and is given by $\Sigma= 13 \times (a/1~\rm{AU})^{-2.35}$.}
\label{fig1}
\end{figure} 

Figures \ref{fig4} and \ref{fig1} show the mass surface density in solids needed to form the observed {\it Kepler} planets {\it in situ} as {\it isolation masses} (i.e., Equation (\ref{e0})), and with a phase of giant impacts (i.e., Equation (\ref{e2})), respectively. The mass surface densities that we find are higher than those calculated in previous works, since these works assumed that solids can be accreted over an annulus with width of order $a$ \citep{CL13}. The best fit disk surface density model of {\it Kepler} planets with $R < 5 R_{\earth}$ is $\Sigma= 13 \times (a/1~\rm{AU})^{-2.35} $. This scaling is steeper than that found by \citet{CL13} because of the additional $a^{1/2}$ dependence on $\Delta a$ in Equation (\ref{e3}).

\subsection{Disk Stability}
The Toomre instability criterion for a gas disk is
\begin{equation}
Q_{Gas}\equiv \frac{c_s \Omega}{\pi G \Sigma_{gas}} < 1
\end{equation}
\citep{T64,GL65}.
Assuming an isothermal disk with a temperature of $10^3$ K and a gas-to-dust ratio of $\Sigma_{gas}/\Sigma=200$ \citep{DA01} yields
\begin{equation}
Q_{Gas} \simeq 4 \times \left(\frac{a}{0.1~\rm{AU}}\right)^{-3/2} \left(\frac{\Sigma}{10^4~\rm{g~cm^{-2}}}\right)^{-1}.
\end{equation}
The upper and lower solid red lines  in Figures \ref{fig4} and \ref{fig1} show the Toomre $Q_{Gas}$ stability parameter $\sim 1$ for the corresponding gas disk with a gas-to-dust ratio of 200 assuming a planet formation efficiency of 100\% and 20\%, respectively. A planet formation efficiency of 100\% means that all the solids in the accretion zone of width $\Delta a$ are ultimately accreted onto the planet, whereas a planet formation efficiency of $\epsilon=20\%$ implies that only one fifth of the solids end up as planets.

Figure \ref{fig4} shows that, even if we assume a 100\% planet formation efficiency, a significant fraction of {\it Kepler} systems fall above the gravitational stability limit, implying that such gas disks are gravitationally unstable to collapse. From this we conclude that these planets therefore cannot have formed as {\it isolation masses} at their current locations.
Figure \ref{fig1} shows that if most close-in {\it Kepler} planets were assembled by giant impacts with a planet formation efficiency of $\epsilon=100\%$, then the corresponding gas disks of {\it Kepler} planets with $R<5 R_{\earth}$ fall close to, but typically below, the gravitational instability limit. If the planet formation efficiency was somewhat less than 100\% then many of the corresponding gas disks would be unstable. Although our findings don't rule out {\it in situ} formation by giant impacts, the initial gas disks would have to have been close to the gravitational stability limit. 

\subsection{Comparison to the Minimum Mass Solar Nebular}
\begin{figure}[htp]
\centerline{
\epsfig{file=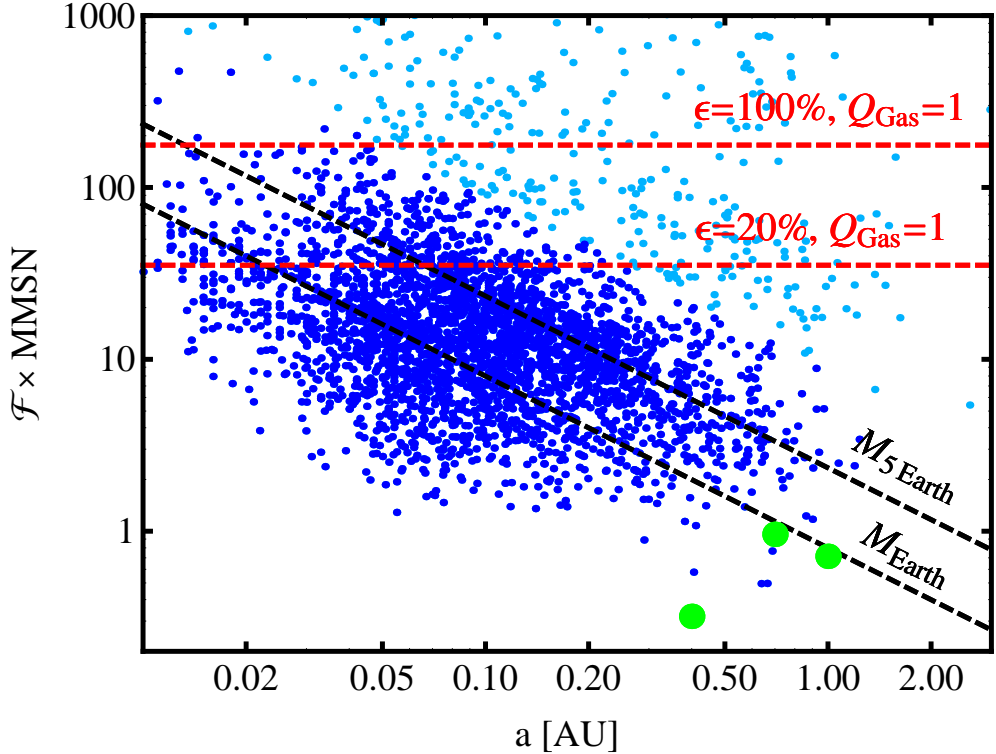, scale=1.3}}
\caption{Enhancement factor above the MMSN, $\mathcal{F}=\Sigma/\Sigma_{MMSN}$, needed for {\it in situ} formation as a function of semi-major axis. Planetary candidates discovered by {\it Kepler} are represented by blue points, where the dark blue points correspond to systems with planetary radii $R \leq 5 R_{\earth}$ and the light blue points to systems with planetary radii $R>5 R_{\earth}$. For comparison, the green points correspond, from right to left, to Earth, Venus and Mercury. The lower and upper dashed-black lines display the enhancement factors needed to form an $1M_{\earth}$-planet and $5M_{\earth}$-planet, respectively. The red dashed lines give the Toomre $Q$ parameter for the corresponding gas disk, $Q_{Gas}$, assuming a gas-to-dust ratio of 200 and a planet formation efficiency of 100\% and 20\%, respectively.}
\label{fig2}
\end{figure}

It is instructive to compare our minimum disk masses for {\it in situ} formation with the MMSN. Normalizing Equation (\ref{e2}) to the MMSN yields
\begin{equation}
\frac{M_{max}}{M_{\earth}}\simeq 1.4 \times \left(\frac{\mathcal{F}a}{1~\rm{AU}}\right)^{3/2}
\end{equation}
where $\mathcal{F}$ is the enhancement factor in solids above the MMSN, $\mathcal{F}\equiv \Sigma/\Sigma_{MMSN}$. This implies that for a MMSN radial disk density profile, the maximum planet mass decreases as $a^{3/2}$. Forming close-in planets {\it in situ} therefore requires a significant enhancement in solids over the MMSN. For example, $\mathcal{F} \sim 20$ and $\mathcal{F} \sim 100$ are required to form a 5~$M_{\earth}$ planet at 0.1~AU and 0.02~AU, respectively. Figure \ref{fig2} shows the enhancement factor needed to form the {\it Kepler} candidates {\it in situ}. Most {\it Kepler} systems require disk masses that are significantly enhanced above the MMSN for {\it in situ} formation. In contrast, formation beyond a few AU is fully consistent with MMSN disk masses.

\section{Conclusions}
We have calculated required disk masses to form close-in super-Earths and mini-Neptunes {\it in situ} from {\it isolation masses}, and find that standard gas-to-dust ratios yield gas disks that are gravitationally unstable for a significant fraction of systems, ruling out such a scenario. In addition, we showed that the maximum width of a planet's accretion region in the absence of any migration is $2 v_{esc}/\Omega$. This maximum width is due to the fact that planets can gravitationally excite their velocity dispersions to values comparable to their escape velocities, but not significantly beyond that. We used this maximum accretion width to calculate the required disk masses for {\it in situ} formation  of the observed {\it Kepler} systems with giant impacts. Our results imply that, even with giant impacts, formation without migration of solids or planets requires typical disk surface densities of solids at semi-major axis less than 0.1~AU of $10^3-10^5 \rm{~g~cm^{-2}}$. This corresponds to typical enhancements above the minimum-mass solar nebular (MMSN) by at least a factor of 20. For standard dust-to-gas ratios this yields gas disk masses close to the gravitational stability limit. These findings are not sensitive to the exact form of the mass-radius relationship. Using published mass-radius relationships \citep{L11,WM14}, instead of simply assuming a density of $2~\rm{g~cm^{-3}}$, strengthens our results somewhat since these relationships yield more massive planets for $R<3R_{\Earth}$, compared to our mass-radius relationship, increasing the values of $\Sigma$ that make up the lower envelopes in Figures 2 and 3, and hence increasing the number of systems that lie close to, or above, the gravitational stability limit. Furthermore, we find that the best fit mass surface density profile for the solids in the disk inferred from the population of {\it Kepler} planets is $\Sigma= 13 \times (a/1~\rm{AU})^{-2.35} $. However, such disk density profiles are much steeper than those inferred from sub-millimeter observations of cold dust in the outer parts of protoplanetary disks, which typically find surface density profiles $\propto a^{-1.0}$ \citep[e.g.][]{A09}. 
This leads us to conclude that, in stark contrast to the terrestrial planets in our Solar System, which likely formed close to their current location from the material locally available in the disk, the formation of close-in super-Earths and mini-Neptunes requires either the transport of large quantities of solids to the inner disk \citep{HM12,CT14}, significantly decreasing the local dust-to-gas ratio, or formation at larger semi-major axis and subsequent migration to their current locations.

Recent sub-millimeter observations \citep{ A12} and theoretical modeling \citep{BA14} suggest that drift in viscous disks rapidly modifies the radial distribution of dust-to-gas ratios in the outer parts of protoplanetary disks such that the standard assumption that $\Sigma_{Gas} \sim 200 \Sigma$ is no longer valid. No such observations exist for the inner most parts of the disk, but it is possible that radial drift gives rise to a significant increase in the amount of solids locally available. Since migration of solids increase the fraction of solids available relative to the gas, it offers a way to locally increase the solid disk surface densities without making the gas disks so massive that they become gravitationally unstable. True {\it in situ} formation is very inefficient at small semi-major axis (see Figure \ref{fig3}) and it should have produced a larger fraction of multiple-planet systems than observed. Even with migration of solids, planet formation efficiencies will remain low, unless material can be trapped locally or most of the solids are accreted by a single growing planet, requiring almost complete accretion as the solids drift through the planet's feeding zone.

Planet formation at larger semi-major axis and subsequent migration offers the other solution for the formation of the observed close-in {\it Kepler} planets. Formation of super-Earths and mini-Neptunes at distance of 1~AU or larger requires no significant enhancement above the MMSN (see Figure \ref{fig2}). For example, a MMSN type disk would be sufficient for the formation of a 5 $M_{\earth}$ planet at 2~AU. The outcome of type I migration, when both migration and eccentricity damping due to the planet's interaction with the gas disk are considered, is consistent with the observation that most ($\gtrsim 90 \%$) {\it Kepler} planets are currently not in or near mean-motion resonances \citep{GS14}. Furthermore, a significant fraction of close-in super-Earths and mini-Neptunes are thought to have large gaseous envelopes containing up to 1\% - 10\% of their total mass. Models examining the accretion and subsequent photo-evaporation of such gaseous envelopes favor formation at a few AU and subsequent inward migration over {\it in situ} formation \citep{LF12,BL14}.

\bibliographystyle{aj} 

\end{document}